# Simulation of the elementary evolution operator with the motional states of an ion in an anharmonic trap

**Ludovic Santos[1], Yves Justum[2], Nathalie Vaeck[1] and M. Desouter-Lecomte[2,3]**

[1] *Laboratoire de Chimie Quantique et Photophysique, CP 160/09 Université Libre de Bruxelles, B-1050 Brussels, Belgium*

[2] *Laboratoire de Chimie Physique, UMR 8000 and CNRS, Université Paris-Sud, F-91405 Orsay, France*

[3] *Département de Chimie, Université de Liège, Bât B6c, Sart Tilman, B-4000 Liège, Belgium*

Following a recent proposal of L. Wang and D. Babikov, J. Chem. Phys. 137, 064301 (2012), we theoretically illustrate the possibility of using the motional states of a $Cd^+$ ion trapped in a slightly anharmonic potential to simulate the single-particle time-dependent Schrödinger equation. The simulated wave packet is discretized on a spatial grid and the grid points are mapped on the ion motional states which define the qubit network. The localization probability at each grid point is obtained from the population in the corresponding motional state. The quantum gate is the elementary evolution operator corresponding to the time-dependent Schrödinger equation of the simulated system. The corresponding matrix can be estimated by any numerical algorithm. The radio-frequency field able to drive this unitary transformation among the qubit states of the ion is obtained by multi-target optimal control theory. The ion is assumed to be cooled in the ground motional state and the preliminary step consists in initializing the qubits with the amplitudes of the initial simulated wave packet. The time evolution of the localization probability at the grids points is then obtained by successive applications of the gate and reading out the motional state population. The gate field is always identical for a given simulated potential, only the field preparing the initial wave packet has to be optimized for different simulations. We check the stability of the simulation against decoherence due to fluctuating electric fields in the trap electrodes by applying dissipative Lindblad dynamics.

## I. INTRODUCTION

Over the last years many works have addressed quantum information processing.[1,2] A quantum computer encodes information in states of quantum bits or qubits, i.e. two-state systems which preserve superpositions of values 0 or 1 unlike classical bits and can manage several tasks in parallel. On a more concrete level, an ideal physical system for computing is isolated from the environment and the states are controlled by fields in order to induce the quantum gate unitary transformations. The coupling with the surroundings is established during the readout process. As predicted by Feynman,[3] quantum computers could be used as quantum simulators to solve stationary[4,5,6,7,8,9] or non stationary[10,11,12,13] quantum problems by simulating them with a controllable experimental setup which allows one to reproduce the dynamics of a given Hamiltonian. Several physical supports have been proposed to encode qubits[14]: photons,[15] spin states using NMR technology,[16] quantum dots,[17] atoms,[18] molecular rovibrational levels of polyatomic or diatomic molecules,[19,20,21,22,23,24,25,26,27,28,29,30,31,32,33,34,35,36,37,38,39,40,41,42,43,44,45,46,47] ultracold polar molecules[48,49,50,51,52,53,54,55,56,57] or a juxtaposition of different types of systems.[58] In the current work we focus on trapped ions[59,60,61,62,63,64] which remain one of the most attractive candidates due to the long coherence timescales and the possibility of exploiting the strong Coulomb interaction.[65] Ultra cold atomic ions with a single outer electron can be trapped in a linear radio frequency Paul trap. Here we consider the $^{111}Cd^+$ ion. The original scheme of information processing using cold trapped ions is to encode the qubit states onto two stable electronic states which can be coupled to the translational states in the trap.[59,64] To improve the fidelity of the gates based on the electronic transitions, it has also been suggested to use an architecture based on an anharmonic quartic trap that could experimentally be realized with a five-segment electrode geometry.[66] Instead of encoding into electronic qubits, it has furthermore been proposed to proceed with the motional states of such an anharmonic trap.[67,68,69] The anharmonicity along the axial direction renders the states energetically non-equidistant and allows one to address the different transitions among the computational basis states. The gates are then driven by electric fields in the radio frequency (*rf*) range.[70] This scheme becomes analog to the control of the vibrational states of a diatomic molecule but in a completely different spectral range. Wang and Babikov[69] have recently numerically simulated a four-qubit Shor's algorithm driven by *rf* pulses obtained by multi-target optimal control theory (MTOCT). In this work, we also use the motional states of an anharmonic ion trap to implement the gate corresponding to the elementary evolution operator of a one-dimensional Schrödinger equation. Numerical algorithms to solve the time-dependent Schrödinger equation (TDSE) usually involve a space and time discretization. The simulator maps the resulting spatial grid points on the selected qubit states.[2,71] In the time domain, an initial proposal focuses on the Split Operator (SO) formalism which involves simple elementary gates as phase shift and quantum Fourier Transform.[71] However, a compromise must be made between the simplicity of the gates and their number so it would be better to use fewer gates and therefore a larger time interval. This strategy has also been adopted in a recent NMR experimental application of a quantum dynamics simulator with three qubits in the case of an isomerization described by a one-dimensional double well.[72,73] The gate matrix could then be calculated by any algorithm, for instance the Chebychev recursion.[74] The present work and the NMR experiment are both Born-Oppenheimer type problems focusing on the nuclear dynamics in a single electronic potential energy curve. The possibility of simulating the full nuclear and electronic dynamics using large scale quantum computers build thanks to trapped ions has also been proposed recently.[10] Finally, the stability of the optimal electric fields driving the elementary evolution is checked by performing dissipative dynamics in order to consider fluctuations in the trap potential due to external fields generated by dipoles in the electrodes.[75,76,77,78,79]

This paper is organized as follows. In the next section, we describe the model of the anharmonic trap and the dissipative approach. The quantum dynamics simulator is described in Sec.III. The optimal control equations are briefly described in Sec. IV. The results are presented in Sec.V and Sec.VI concludes.

## II. ION TRAP MODEL

We consider a single $^{111}$Cd$^+$ ion in a Paul trap. Its axial harmonic frequency is $\nu = \omega/2\pi = 2.77$ MHz.[80,81] Following the proposal of Babikov and coworkers,[67,68,69] the trapping potential is assumed to be slightly anharmonic to break the regular energy spacing between the states. The axial coordinate is denoted $z$ and the field-free Hamiltonian reads

$$H_0 = -\frac{\hbar^2}{2m}\frac{\partial^2}{\partial z^2} + q\left(\frac{k}{2}z^2 + \frac{k'}{4!}z^4\right) = -\frac{\hbar^2}{2m}\frac{\partial^2}{\partial z^2} + qW(z) \tag{1}$$

where $m$ and $q$ are the mass and charge of the ion. The force constant corresponding to a frequency $\nu = 2.77$ MHz is $k = 3.5828 \times 10^{-14}$ a.u. and we choose $k' = 3.5828 \times 10^{-18}$ a.u. to model the anharmonic contribution. This value for the anharmonicity $k'$ is slightly larger than the one considered in the Zhao Babikov model. It has been chosen to ensure convergence of the dynamical basis set which contains 32 eigenstates. However, our simulation involves only 16 states. The eigenstates are obtained by diagonalizing the Hamiltonian $H_0$ in a basis set of 50 harmonic oscillator eigenfunctions.[82] The control of the dynamics is carried out by coupling the system with an electric field assumed to be independent of the axial coordinate $z$. The Hamiltonian is then:

$$H(t) = H_0 - qzE(t) \qquad (2)$$

and its evolution is described by the TDSE

$$i\hbar \frac{\partial}{\partial t} \Phi(z,t) = H(t)\Phi(z,t) \qquad (3)$$

where $\Phi(z,t)$ is the wave function of the ion. In the dissipative dynamics, we replace the wave function by the matrix $\rho(t)$ of the density operator:

$$\hat{\rho}(t) = \left| \Phi(t) \right\rangle \left\langle \Phi(t) \right|. \qquad (4)$$

In order to take into account the perturbation due to fluctuating fields on the trap electrodes,[75,76,77,78,79] we performed simulations following the Lindblad formalism[83,84,85] in which the system density matrix evolves according to the master equation:

$$\frac{\partial \rho(t)}{\partial t} = -\frac{i}{\hbar} \left[ H(t), \rho(t) \right] + L_D \rho(t) \qquad (5)$$

with

$$L_D \rho(t) = \sum_{jk} \left( L_{jk} \rho(t) L_{jk}^\dagger - \frac{1}{2} \left[ \rho(t), L_{jk}^\dagger L_{jk} \right]_+ \right) \qquad (6)$$

where $[.,.]_+$ denotes the anticommutator. The Lindblad operators $L_{jk}$ are written phenomenologically as transition operators among the eigenstates of the field-free Hamitonian $H_0$:[85]

$$L_{jk} = \sqrt{\gamma_{jk}}\,|j\rangle\langle k| \qquad (7)$$

and $\gamma_{jk}$ is the transition rate. Since the coupling is due to external fields generated by fluctuating dipoles in the trap electrodes, we assume that the rate depends on the dipole transition moments $\mu_{jk} = q\langle j|z|k\rangle$ and we set:

$$\gamma_{jk} = \kappa\,|\mu_{jk}|. \qquad (8)$$

This $\kappa$ parameter has been calibrated to still obtain good results for the simulation despite the dissipation and this allows us to estimate the relevant minimum average heating time $\bar{\gamma}^{-1}$ for a given pulse duration and a given simulation.

## III. QUANTUM DYNAMICS SIMULATION

To illustrate the quantum dynamics simulator, we consider a one-dimensional model for a particle of mass $m_s$ in an arbitrary potential $V(x)$. The Hamiltonian of the simulated system is denoted $H_s$ while the field-free Hamiltonian of the simulator is $H_0$ [Eq.(1)]. One has:

$$H_s = -\frac{\hbar^2}{2m_s}\frac{\partial^2}{\partial x^2} + V(x) \qquad (9)$$

The aim of the quantum simulator is to obtain the wave function $\psi(x,t)$ starting from any initial condition $\psi(x,t=0)$ for a given potential $V(x)$. The numerical algorithms implemented on classical computers usually discretize the spatial coordinate $x$ and the time $t$. In Cartesian coordinates, $\psi(x,t)$ is described on an equally-spaced grid with an interval $\Delta_x = L/N$ where $L = x_{max} - x_{min}$ is the grid length and $N$ the number of points. Each grid point is given by:

$$x_j = x_{min} + (j+1)\Delta_x . \tag{10}$$

where $j = 0,..., N-1$. In the time domain, the propagation is divided into $N_p$ steps such that

$$t_{propagation} = N_p \Delta t . \tag{11}$$

When the potential energy does not depend on time, the final wave function $\psi(x, t_{propagation})$ can be obtained by the iterative application of the elementary evolution operator $U_s(\Delta t)$ starting from the initial condition

$$\psi(x, N_p\Delta t) = U_s(\Delta t)....U_s(\Delta t)\psi(x,t=0) . \tag{12}$$

On a grid with $N$ points, $U_s(\Delta t)$ is a $N$ x $N$ unitary matrix which defines the gate of the simulator. Experimentally, it is more interesting to reduce the number of pulses and thus to choose a relatively large $\Delta t$ in order to perform the simulation involving $N_p\Delta t$ with a small $N_p$. Some algorithms such as the Chebychev recursion[74] are adapted to large time intervals. However, we choose here the Split Operator algorithm**Erreur ! Signet non défini.**[86] which requires a small time step $\delta t$ to be accurate up to terms of order $\delta t^3$. As the chosen time interval $\Delta t$ is too large to directly express $U_s(\Delta t)$ following the Split Operator algorithm with a small error, we divided it in $K$ smaller time intervals $\delta t = \Delta t/K$. The transformation matrix is then given by:

$$U_s(\Delta t) = U_s(K\delta t) = (U_s(\delta t))^K . \tag{13}$$

Now, we can calculate $U_s(\delta t)$ by:

$$U_s(\delta t) = e^{-\frac{i}{\hbar}V\frac{\delta t}{2}} QFT^\dagger e^{-\frac{i}{\hbar}T\delta t} QFT e^{-\frac{i}{\hbar}V\frac{\delta t}{2}} \qquad (14)$$

As already suggested,[2,10,71,72,73,11] the Split Operator algorithm is particularly suited for inducing the elementary evolution by simple gates such as quantum Fourier transform $(QFT)$[1] and controlled phase-shift gates. The $QFT$ gate changes the position representation to the momentum representation. The exponential operators are diagonal in the basis sets in which they act so that the transformation consists only in modifying the phase of each basis state. The phases are changed simultaneously for all the states belonging to the evolving superposition. It is interesting to note that an algorithm based on the Walsh functions[87] has recently been proposed to implement diagonal unitary transformation.[88] In principle, the resources needed to compute using a SO scheme should depend on the number of time steps and grid points in a polynomial way, while the corresponding classical resources would increase exponentially. However, the decomposition of the SO transformation into five elementary gates may cause the accumulation of experimental errors and increase decoherence as already discussed in the experiment using NMR technology.[72,73] As suggested in this work, it is more efficient to directly optimize the field steering $U_s(\Delta t)$ and by doing so, to perform the optimization of a "black box" for a given time interval. The corresponding unitary matrix only depends on the potential and the particle mass but remains the same for the simulation of any initial condition $\psi(x, t = 0)$.

The wave function $\psi(x, t = 0)$ discretized on the $N$ grid points is encoded in $N$ states of the quantum simulator. Each qubit basis state corresponds here to a motional eigenvector $\chi_j(z)$ of the Hamiltonian $H_0$ [Eq.(1)]. This means that $\psi(x, t = 0)$ is mapped on the wave function $\Phi(z, t = 0)$ of the ion expressed in the eigen basis set:

$$\Phi(z, t = 0) = \sum_{j=0}^{N-1} c_j(t=0) \chi_j(z) \qquad (15)$$

The localization probabilities $\left|\psi(x_j, t=0)\right|^2$ at grid point $x_j$ is thus mapped on the population $\left|c_j(t)\right|^2$ in the qubit state by taking into account the normalization condition

$$\left|\psi(x_j, t=0)\right|^2 \leftrightarrow \left|c_j(t=0)\right|^2 / \Delta_x \qquad (16)$$

The *rf* field is optimized to induce the unitary transformation $U_s(\Delta t)$ among the eigenstates of the ion at the end of the pulse where $t = t_{pulse}$. The pulse duration is mainly determined by energy spacing in the selected computational basis set. The simulator wave packet evolves by successive applications of the evolution operator $U(t_{pulse})$ containing the gate field $E(t)$

$$\Phi(z, N_p t_{pulse}) = U(t_{pulse})....U(t_{pulse}) \Phi(z, t=0) \qquad (17)$$

The total duration of the simulation with the time dependent Hamiltonian $H(t)$ [Eq.(2)] is thus

$$t_{simulation} = N_p t_{pulse} \qquad (18)$$

and the localization probability can be obtained after the $l^{\text{th}}$ pulse by the relation:

$$\left|\psi(x_j, l\Delta t)\right|^2 \leftrightarrow \left|c_j(t = l t_{pulse})\right|^2 / \Delta_x \qquad (19)$$

## IV. OPTIMAL FIELD DESIGN

The target unitary transformation is now denoted $U_s \equiv U_s(\Delta t)$ [Eqs.(13) and (14)] where we dropped the chosen time interval. The optimal field $E(t)$ must induce the transformation among the qubit states after the time $t_{pulse}$. The field $E(t)$ is designed by the multi-target optimal control theory (MTOCT).[20] At the end of $l^{th}$ pulse, the amplitudes of the basis states must be $\mathbf{c}(lt_{pulse}) = U_s \mathbf{c}((l-1)t_{pulse})$ at an arbitrary phase which must be the same for all the transitions. The evolution operator of the simulator with the optimum field is $U(t_{pulse})$. The gate performance must then measure by a phase sensitive quantity. We use the fidelity[89,90] built from the overlap between each target state $U_S |j\rangle$ and the corresponding final state obtained by the control $U_P(t_{pulse}) |j\rangle$ where $U_P(t_{pulse}) = PU(t_{pulse})P$ and $P$ projects on the $N$ states

$$F = \left| \sum_{j=1}^{N} \langle U_S j | U_P(t_{pulse}) j \rangle \right|^2 / N^2 = \left| Tr\left( U_s^\dagger U_P(t_{pulse}) \right) \right|^2 / N^2 \qquad . \qquad (20)$$

The optimal field is obtained by maximizing a functional based on an objective and constraints to limit the total integrated intensity and to ensure that the Schrödinger equation is satisfied during the process.[91] Several possibilities differing by the choice of the objective have been discussed in the literature. A first proposal is based on the fidelity [Eq.(14)][89,90,44,92]

$$\begin{aligned}
J_F = {} & \left| Tr\left( U_s^\dagger U_P(t_{pulse}) \right) \right|^2 - \int_0^{t_{pulse}} \alpha(t) E^2(t) dt \\
& - 2\Re e\left[ \sum_{j=1}^{N} \langle j | U_s^\dagger U_P(t_{pulse}) | j \rangle \sum_{k=1}^{N} \int_0^{t_{pulse}} \langle \lambda_k(t) | \partial_t + \frac{i}{\hbar} H(t) | k(t) \rangle dt \right]
\end{aligned} \qquad (21)$$

where $\alpha(t) = \alpha_0 / \sin^2(\pi t / t_{pulse})$ and $\lambda_k(t)$ is the Lagrange multiplier for the Schrödinger equation constraint. The $|j\rangle$ states are the qubit basis state. An other strategy is to use the sum of transition probabilities between each basis state $|j\rangle$ and the target $U_s |j\rangle$ [24,31]

$$J_P = \sum_{j=1}^{N+1} \left| \left\langle j \left| U_s^\dagger U_P(t_{pulse}) \right| j \right\rangle \right|^2 - \int_0^{t_{pulse}} \alpha(t) E^2(t) dt$$
$$-2\Re e \left[ \sum_{j=1}^{N+1} \left\langle j \left| U_s^\dagger U_P(t_{pulse}) \right| j \right\rangle \int_0^{t_{pulse}} \left\langle \lambda_j(t) \left| \frac{\partial}{\partial t} + \frac{i}{\hbar} H(t) \right| j(t) \right\rangle dt \right] \qquad (22)$$

where a supplementary transition involving an initial superposed state has to be added in order to ensure a good phase control so that the gate is valid for any superposition

$$2^{-N/2} \sum_{j=1}^{N} \left| j \right\rangle \rightarrow 2^{-N/2} \sum_{j=1}^{N} U_s \left| j \right\rangle e^{i\phi} \qquad (23)$$

where $\phi$ is a single phase taking any value between 0 and $2\pi$. Variation with respect to $\left\langle \lambda_j \right|$ leads to an evolution equation with an initial condition $\left| j \right\rangle$ and the variation of $\left| j \right\rangle$ gives an equation with the target state $U_s \left| j \right\rangle$ as a final condition. Depending on the strategy, $N$ or $N + 1$ forward and backward propagations are thus required. The field is built from a contribution of all the wave packets and is given by:

$$E_F(t) = -\frac{1}{\alpha(t)} \Im m \left[ \sum_{j=1}^{N} \left\langle \lambda_j(t) \left| j(t) \right\rangle \sum_{k=1}^{N} \left\langle \lambda_k(t) \left| \mu \right| k(t) \right\rangle \right. \right]. \qquad (24)$$

for the functional $J_F$ [Eq.(21)] and by

$$E_P(t) = -\frac{1}{\alpha(t)} \Im m \left[ \sum_{j=1}^{N+1} \left\langle \lambda_j(t) \left| j(t) \right\rangle \left\langle \lambda_j(t) \left| \mu \right| j(t) \right\rangle \right. \right] \qquad (25)$$

for the functional $J_P$ [Eq.(22)]. The MTOCT equations are solved by the Rabitz iterative monotonous convergent algorithm.[91] At each iteration step $i$ the field is obtained by $E^{(i)} = E^{(i-1)} + \Delta E^{(i)}$ where $\Delta E^{(i)}$ is estimated from Eq. (24) or Eq.(25).

To include decoherence effects, we use the extension of the monotonically convergent algorithm to treat the system with dissipation.[93] The wave functions $\left| j(t) \right\rangle$ and $\left| \lambda_j(t) \right\rangle$ are replaced by the density matrices $\rho_j(t)$ and $\eta_j(t)$ respectively. In the superoperator notation, one writes $\left|\rho_j(t)\right\rangle\rangle$, $\left|\eta_j(t)\right\rangle\rangle$ and the scalar product becomes $\left\langle\langle\eta_j(t)\big|\rho_j(t)\right\rangle\rangle = Tr\left(\eta_j^\dagger(t)\rho_j(t)\right)$. The forward and backward propagations are carried out using the Lindblad master equation [Eq.(4)]. The expression of the field becomes

$$E_F(t) = -\frac{1}{\alpha(t)}\Im m\left[\sum_{j=1}^N\left\langle\langle\eta_j(t)\big|\rho_j(t)\right\rangle\rangle\sum_{k=1}^N\left\langle\langle\eta_k(t)\big|\mathrm{M}\big|\rho_k(t)\right\rangle\rangle\right]$$ (26)

and

$$E_P(t) = -\frac{1}{\alpha(t)}\Im m\left[\sum_{j=1}^{2^n+1}\left\langle\langle\eta_j(t)\big|\rho_j(t)\right\rangle\rangle\left\langle\langle\eta_j(t)\big|\mathrm{M}\big|\rho_j(t)\right\rangle\rangle\right]$$ (27)

where in the superoperator notation $\mathrm{M}\big|\rho_j(t)\rangle\rangle = \big|\mu\rho_j(t)\rangle\rangle - \big|\rho_j(t)\mu\rangle\rangle$.

# V. RESULTS

In the present application we choose a very simple potential to ensure correct dynamics with only 16 grid points. The qubit basis states consist of the 16 lowest motional eigenstates of the trap and we adopt a dynamical basis containing 32 eigenstates. Convergence in this basis set has been checked. The parameters of the simulated problem are $m_s = 1$ a.u. and $V(x) = x^2/2$, i.e. a harmonic potential with frequency $\omega = 1$ a.u. The corresponding period is $2\pi$ a.u. Figures 1 (a) and (c) show the anharmonic potential $W(z)$ of the ion [Eq.(1)] and the simulated harmonic potential $V(x)$ [Eq.(9)]. Figure 1 illustrates the mapping between the simulated wave function at the grid points and the amplitude of the qubit states, i.e. in the motional eigenstates of the ion.

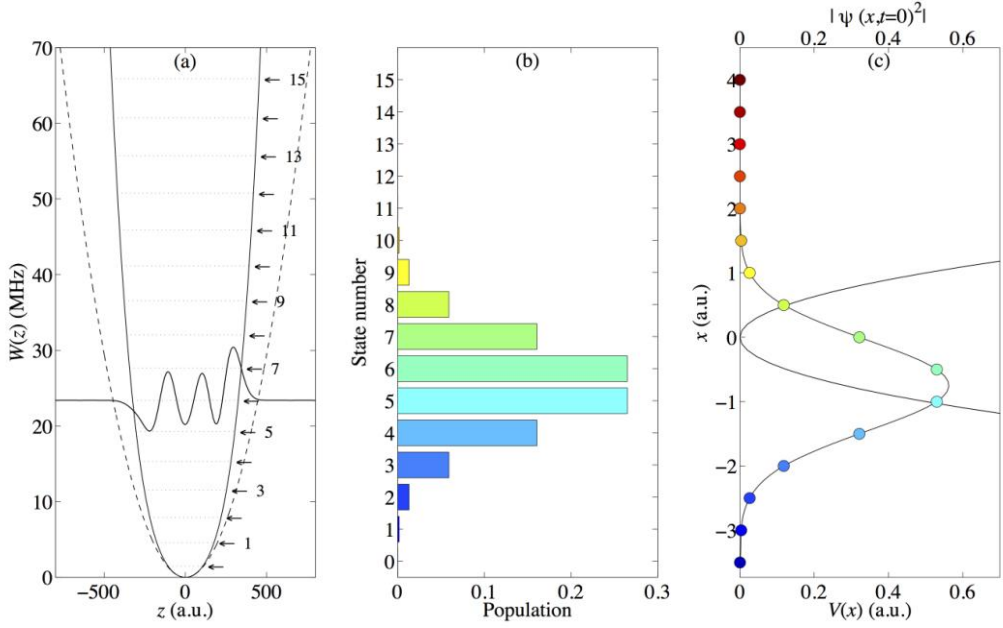

FIG. 1. Mapping in the TDSE simulator. Panel (a): Harmonic (dashes) and anharmonic (full line) potentials of the $Cd^+$ ion trap and the superposed state $\Phi(z, t = 0)$ with initial amplitudes $c_j(t = 0) = \sqrt{\Delta_x} \psi(x_j, t = 0)$ in the $j^{th}$ motional eigenstate of the ion. The eigenenergies are indicated by arrows. Panel (b): Population $\left| c_j(t = 0) \right|^2$ in the qubit states $\left| j \right\rangle$. Panel (c): Simulated system with a harmonic potential $V(x)$ and initial localization probability $\left| \psi(x_j, t = 0) \right|^2$.

We simulate the propagation of Gaussian wave packets

$$\psi(x, t = 0) = (\sigma / \pi)^{1/4} \exp\left[ -(x - x_0)^2 / 2\sigma \right].$$ (28)

The spatial grid extends from $x_{min}$ = - 4 a.u. to $x_{max}$ = 4 a.u. and the 16 grid points are given by Eq.(10). We want to simulate a complete oscillation of the Gaussian wave packet with $N_p$ = 10 $rf$ spots. This means that the unitary transformation corresponds here to $U_s(\Delta t)$ with $\Delta t = 2\pi / 10$ a.u. As it has been explained in section III, we have divided this time interval by $K$ to ensure a good accuracy of the Split Operator. In consequence, we numerically built the matrix $U_s(\Delta t) = (U_s(\delta t))^K$ with $\delta t = 2\pi / 100$ a.u. and $K = 10$. The field driving the

gate $U_s(\Delta t)$ is obtained by MTOCT with a guess field composed of 28 frequencies that

correspond to 15 transitions with a variation of the motional quantum number $\Delta v = \pm 1$ and

13 transitions with $\Delta v = \pm 3$. The initial amplitude is fixed to $1.945 \times 10^{-13}$ a.u., i.e. 0.1 $Vm^{-1}$,

for each frequency. The duration of the $rf$ field is $t_{pulse} = 96$ $\mu s$ and the time step for the propagation is 960 $ps$. The penalty factor is $\alpha_0 = 10^{15}$ a.u. for $J_P$ and $\alpha_0 = 4 \times 10^{15}$ a.u. for

$J_F$. The fields that do not include dissipation have been optimized up to a fidelity of 0.99999.

With such a high fidelity, the results of the simulation are similar whether we apply the field

$E_F(t)$ [Eq.(24)] or $E_P(t)$ [Eq.(25)]. Convergence requires about 1500 iterations. The

propagation of the dynamical equations is carried out in the interaction representation by the

fourth order Runge-Kutta method.[94]

## A. Simulation without dissipation

We first simulate the propagation of a coherent Gaussian wave packet, i.e. the ground vibrational state with $\sigma = \hbar / m_s \omega = 1$ a.u. and an equilibrium position $x_0 = -0.75$ a.u. (see Fig.

1). The ion is assumed to be in the ground motional state $\chi_0(z)$ and we optimize a field for the

initialization of the propagation at $t = 0$. The target is then the ionic initial wave packet

$\Phi(z, t = 0) = \sum_{j=0}^{N-1} c_j(t = 0)\chi_j(z)$ for which the amplitudes are given by

$c_j(t = 0) = \psi(x_j, t = 0)\sqrt{\Delta_x}$. The square modulus of this initial simulator wave packet is the

bold red curve in Fig. 2. The evolution of $|\Phi(z, t)|^2$ after successive applications of

$U(t_{pulse})$ with the field $E_P(t)$ [Eq.(25)] is shown in Fig.2. The fidelity being 0.99999,

similar results are obtained with $E_F(t)$ [Eq.(24)]. The $rf$ pulse drives the $U_s$ transformation

at the end of the pulse. One observes the expected periodicity

$\left|\Phi\left(z,t=(0+l)t_f\right)\right|^2 = \left|\Phi\left(z,t=(10-l)t_f\right)\right|^2$ with $l = 0,\ldots, 4$ in agreement with the corresponding periodic dynamics of the simulated coherent wave packet.

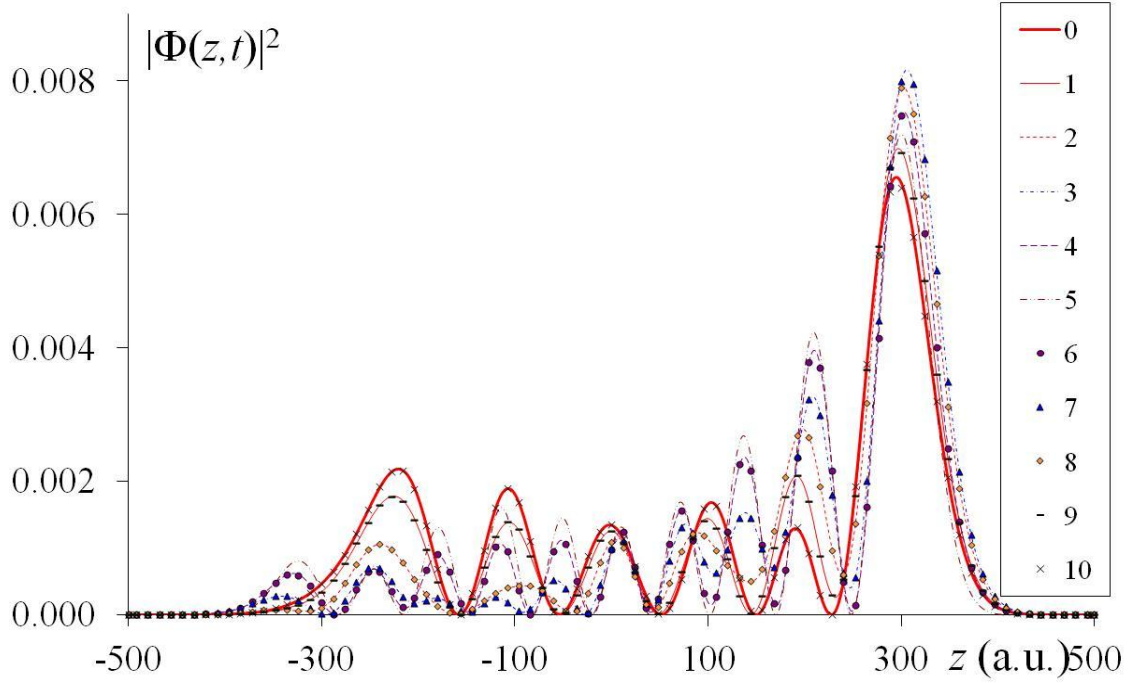

FIG. 2. Evolution of the square modulus of the motional wave function $\left|\Phi(z,t)\right|^2$ after successive applications ($l = 1,..,\ 10$) of the *rf* pulse driving the $U_s$ gate for the simulation of a coherent Gaussian wave packet in a harmonic potential (see Fig. 3a). The legend gives the pulse number. The wave function for $t = 0$ (red thick curve) is prepared by the initialization pulse so that $c_j(t=0)/\sqrt{\Delta_x} = \psi(x_j, t=0)$. The applied field is $E_P(t)$ [Eq.(25)].

Figure 3 compares the results of the simulation (discrete markers) with the exact evolution of the simulated wave packet $\psi(x,t)$ (continuous lines) for $t = l\Delta t$ and $l = 0,\ldots,10$. The eigenstate populations of the simulator after the $l^{\text{th}}$ pulse are mapped on the localization probability of the simulated system $\left|c_k(t=l\Delta t)\right|^2/\Delta_x \leftrightarrow \left|\psi(x_k,t=l\Delta t)\right|^2$. Fig. 3 (a) shows the evolution of the coherent state with $\sigma = 1$ a.u. and with conservation of the initial shape. Fig. 3(b) gives the breathing of a Gaussian packet with $\sigma = 0.5$ a.u. This illustrates that the optimal field executes the transformation $U_s(\Delta t)$ for any initial wave packet. Only the initialization step requires a new optimization.

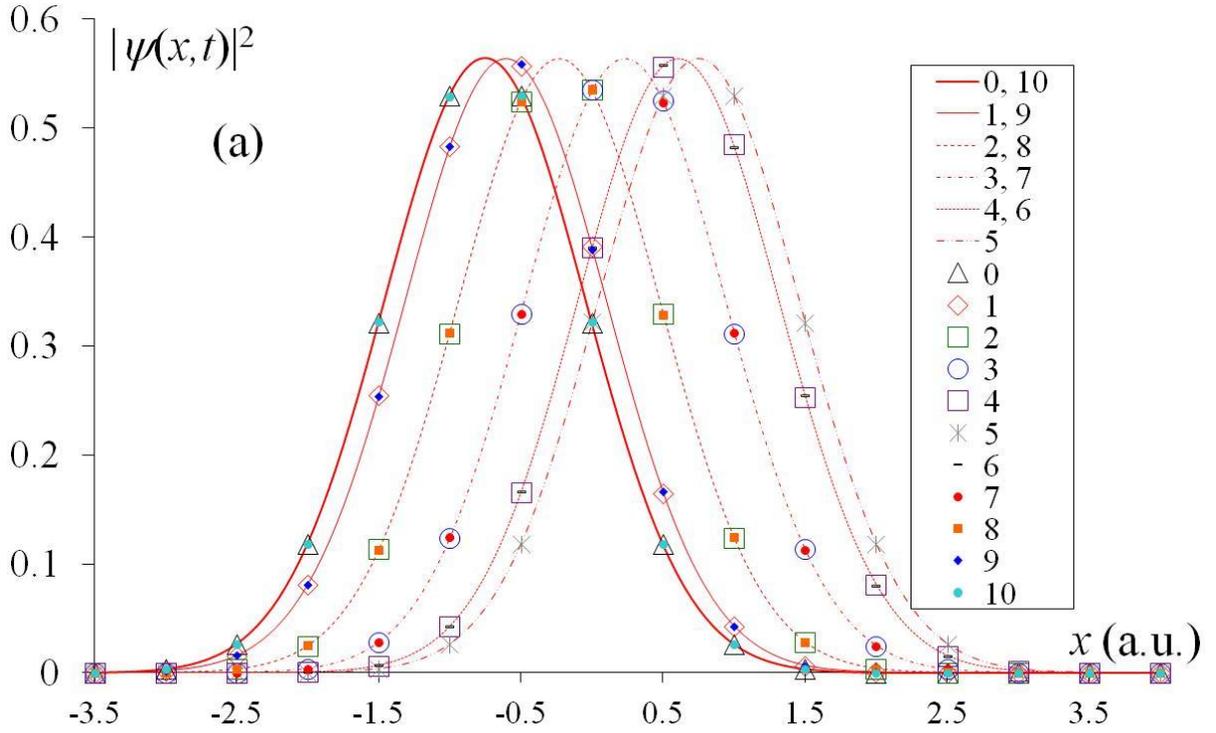

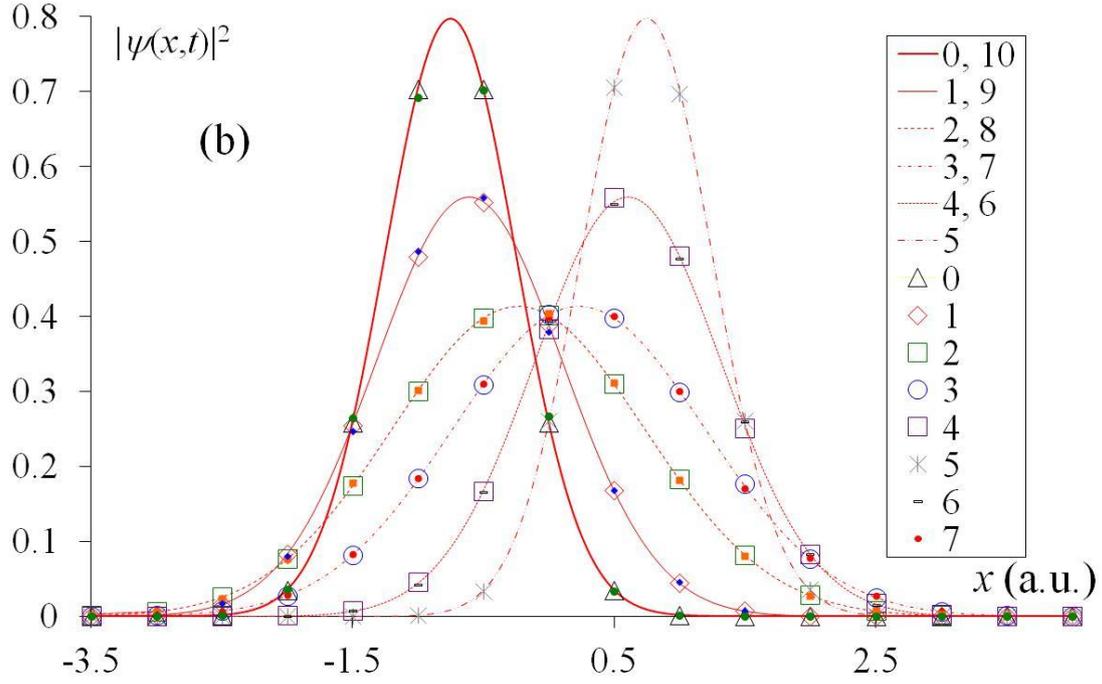

FIG. 3. Exact evolution of Gaussian wave packets (continuous lines) and results obtained from the mapping $\left|\psi(x_j,t)\right|^2 \leftrightarrow \left|c_j(t)\right|^2 / \Delta_x$ with the populations of the ion eigenstates (markers) after application of the $l^{\text{th}}$ pulse. The legend gives the number $l$ of the pulse. One observes the expected periodicity $\left|\psi\left(x, t = (0 + l)\Delta t\right)\right|^2 = \left|\psi\left(x, t = (10 - l)\Delta t\right)\right|^2$. Panel (a): coherent wave packet with $\sigma = 1$ a.u.; panel (b): $\sigma = 0.5$ a.u. The applied field is $E_p(t)$ [Eq.(25)].

Optimum fields and their Fourier transform $\left|S(\nu)\right|^2$ in arbitrary units are shown in Fig. 4. The field for the preparation of the ion wave packet $\Phi(z, t = 0)$ simulating the coherent Gaussian case ($\sigma = 1$ a.u.) is displayed in Fig. 4 (a) and its Fourier transform in Fig. 4 (b) (this case is illustrated in Fig.1). Figs. 4 (c) and (d) show the gate optimum field $E_F(t)$ [Eq.(24)] and its spectrum. The background at high frequencies has been filtered. The fidelity decreases from 0.99999 to 0.97, but a new optimization allows us to reach 0.99999 again. The new $E_F(t)$ field after filtering and reoptimization is shown in Fig. 4 (e) and its spectrum in Fig. 4 (f). One observes that the control has found a new mechanism involving mainly the high frequencies and the noise remains very small. The optimization with the other functional $E_p(t)$ [Eq.(25)] is shown in Fig. 4 (g). One sees in Fig. 4 (h) that this procedure directly gives a simple spectrum and thus does not require a filtering. The maximum amplitude of the field is larger than that of the guess field, but remains acceptable since it does not exceed 1.5 $Vm^{-1}$. The Fourier transform remains very simple. We find the transitions of the guess field with slightly different intensities (they were equal in the trial field). The frequencies correspond to the transition $\Delta \nu = \pm 1$ below 5 MHz and $\Delta \nu = \pm 3$ above 10 MHz. An increasing energy gap is observed due to the anharmonicity of the symmetrical potential (see Fig.1 for the lowest states).

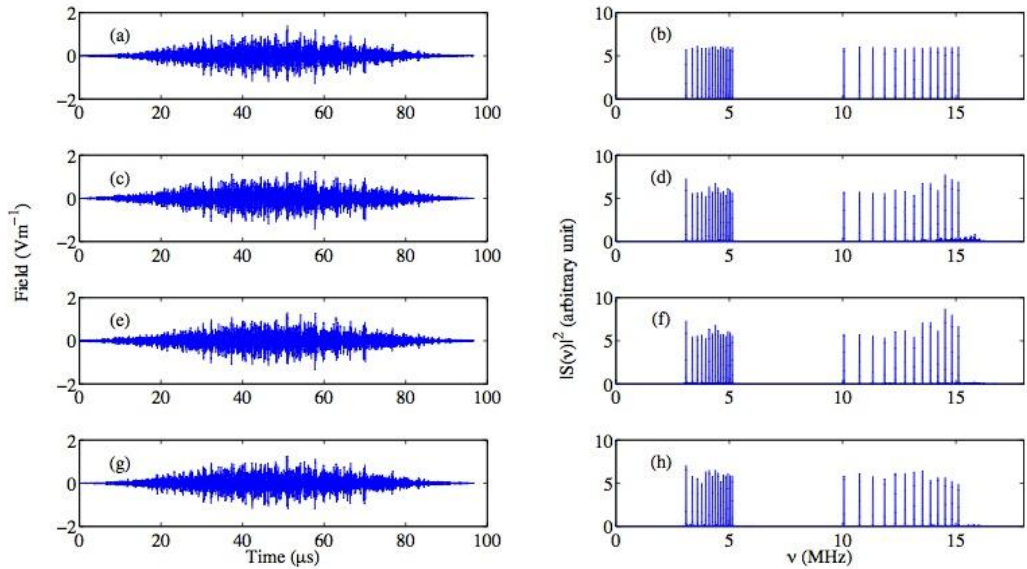

FIG. 4 Optimum fields and their spectrum $|S(\nu)|^2$ in arbitrary units for the preparation step and the simulation of the elementary transformation $U_s(\Delta t)$. Panels (a) and (b): Preparation

of the initial wave packet $\Phi(z, t = 0)$ for the simulation of the coherent Gaussian ($\sigma = 1$ a.u.); panels (c) and (d): field $E_F(t)$ [Eq.(24)], panels (e) and (f): field $E_F(t)$ after filtering of the background and reoptimization, panels (g) and (h): field $E_P(t)$ [Eq.(25)].

## B. Simulation with dissipation

The stability of the simulator against decoherence is now examined by performing Lindblad dissipative dynamics with the fields presented in Fig. 4. We aim at estimating the order of magnitude of the mean heating rate while preserving relevant information about the simulated dynamics. Fig. 5 illustrates the results for a strong dissipation $\kappa = 5 \times 10^{-18}$ a.u. [Eq.(8)] leading to a mean characteristic heating time $\bar{\gamma}^{-1} = 55$ $ms$ while the propagation time is about $1ms$. The average rate $\bar{\gamma}$ is taken over all the transitions with $\Delta\nu = \pm 1, \pm 3$. An increasing discrepancy is observed between the expected wave packet and the simulated points, after the application of a number of $rf$ pulses. In particular, the periodicity $|\psi(x, t = (0 + l)\Delta t)|^2 = |\psi(x, t = (10 - l)\Delta t)|^2$ is not strictly respected anymore. However, the qualitative behavior of the wave packet is still reasonably described.

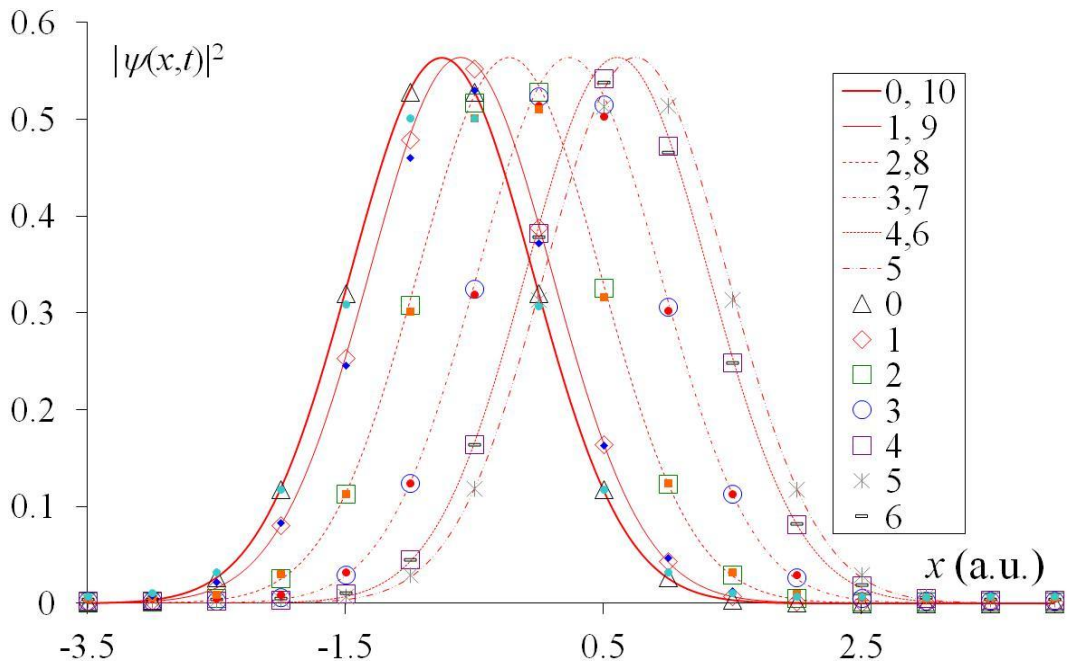

FIG. 5. Simulation in presence of dissipation due to fluctuating electric fields. Exact evolution of the coherent wave packet ($\sigma = 1$ a.u.) (continuous lines) and results obtained from the populations of the ion eigenstates (markers) during the Lindblad dynamics with $\kappa = 5 \times 10^{-18}$ a.u. [Eq.(8)] ($\bar{\gamma}^{-1} = 55$ *ms*) after application of the $l^{\text{th}}$ pulse. The legend gives the number $l$ of the pulse. The applied field is $E_p(t)$ [Eq.(25)].

The evolution of the fidelity [Eq.(20)] during the simulation is given in Fig. 6 for three values of the decoherence strength $\kappa$ [Eq.(8)]: $\kappa = 10^{-17}$ a.u. or $\bar{\gamma}^{-1} = 11$ *ms*, $\kappa = 5 \times 10^{-18}$ a.u. or $\bar{\gamma}^{-1} = 55$ *ms* and $\kappa = 10^{-18}$ a.u. or $\bar{\gamma}^{-1} = 110$ *ms*. The gate field is calculated using $E_F(t)$ [Eq.(24)] (dashed lines) or $E_p(t)$ [Eq.(25)] (full lines). Note that the two $E_F(t)$ fields without (Fig.4 (c)) and with filtering and reoptimization (Fig.4 (e)) give the same results up to the fourth significant digit. The initial fidelity is that after the preparation step which is carried out also with dissipation. One observes that the $E_F(t)$ field is slightly more sensitive to decoherence than $E_p(t)$. This is due to the mechanism induced by $E_F(t)$ which involves higher frequencies transitions. The parameter $\kappa$ must obviously remain smaller than $10^{-18}$ a.u. and thus $\bar{\gamma}^{-1} > 110$ *ms*, to maintain a very high fidelity for a propagation of 1*ms*. This corresponds to the usual expected decoherence time in an ion trap.[14]

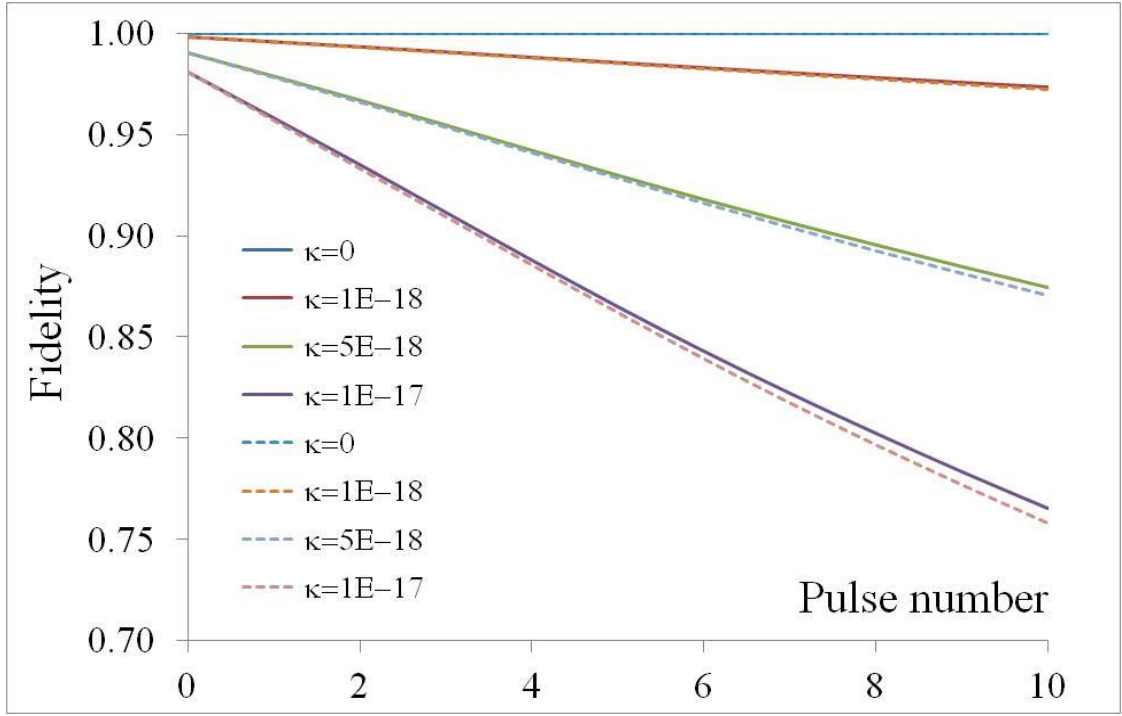

FIG. 6. Evolution of the fidelity [Eq.(20)] during the simulation of the coherent wave packet ($\sigma$ = 1 a.u.) for different values of the decoherence strength [Eq.(8)] with the gate field calculated by $E_F(t)$ [Eq.(24)] (dashed lines) or $E_P(t)$ [Eq.(25)] (full lines). The characteristic decoherence lifetimes are: $\overline{\gamma}^{-1}$ = 11 $ms$ ($\kappa = 10^{-17}$), $\overline{\gamma}^{-1}$ = 55 $ms$, ($\kappa = 5 \times 10^{-18}$) and $\overline{\gamma}^{-1}$ = 110 $ms$ ($\kappa = 10^{-18}$).

Fig. 7 shows the evolution of the mean position of the ion $\left\langle z(t) \right\rangle$ (Fig. 7 (a)) and of the simulated coherent wave packet $\left\langle x(t) \right\rangle$ (Fig. 7 (b)) without dissipation (blue full lines) and with different heating rates (dashed lines) after the successive applications of the gate field calculated by $E_P(t)$ [Eq.(25)] (full lines). During the initialization step, the average position $\left\langle z(t) \right\rangle$ increases from the equilibrium position ($z = 0$) of the trap to a value of $z$ = 114.8 a.u. (6.074 $nm$). During the successive interactions with the *rf* pulses, the periodic behavior characteristic of the decoherence-free case (blue full curve in Fig.7 (a)) is more and more altered, but the qualitative behavior remains acceptable. The aftereffect of the decoherence on the average position of the simulated coherent wave packet is shown in Fig.7 (b). The error increases from 1% after the first pulse to 16% at the end of the simulation for the most dissipative case $\kappa = 10^{-17}$ a.u. ($\bar{\gamma}^{-1} = 11\ ms$).

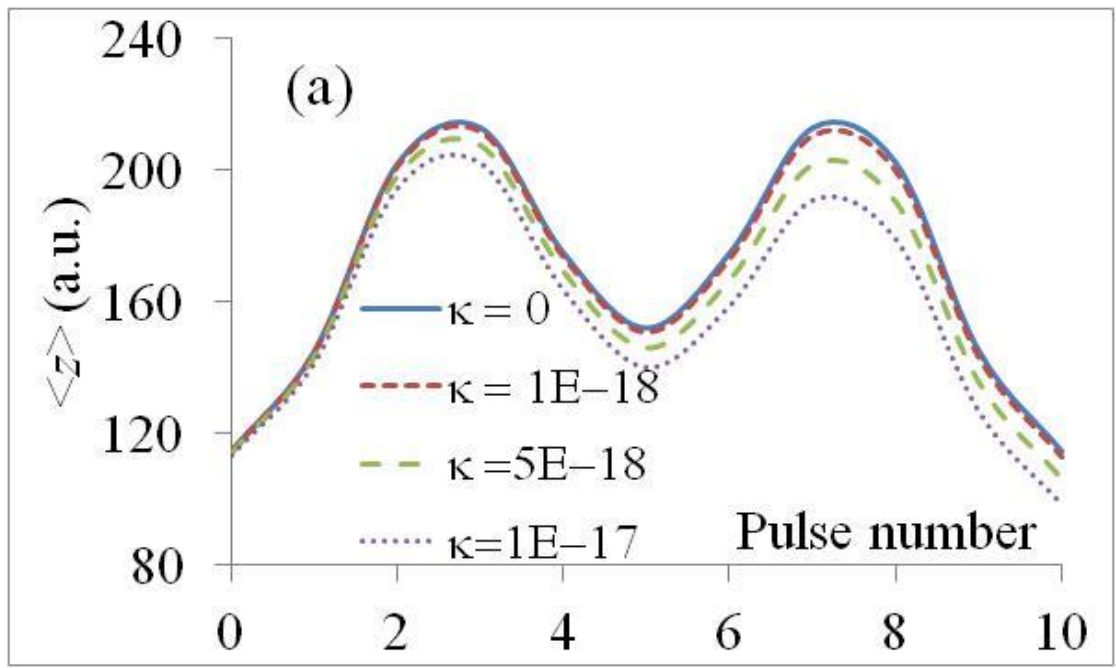

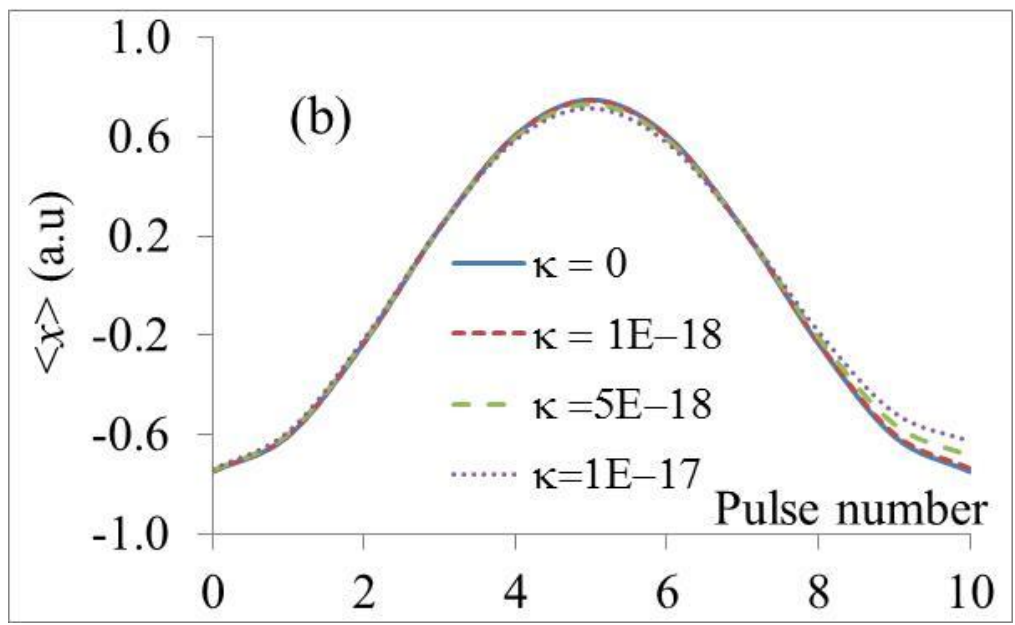

FIG. 7. Evolution of the mean position of the ion $\langle z(t) \rangle$ (panel a) [Eq.(3)] and of the mean position of the coherent wave packet $\langle x(t) \rangle$ (panel b) without decoherence ($\kappa = 0$) and for different values of the decoherence strength [Eq.(8)]. The field is calculated using $E_p(t)$ [Eq.(25)]. The characteristic decoherence lifetimes are: $\bar{\gamma}^{-1} = 11$ $ms$ ($\kappa = 10^{-17}$), $\bar{\gamma}^{-1} = 55$ $ms$, ($\kappa = 5 \times 10^{-18}$) and $\bar{\gamma}^{-1} = 110$ $ms$ ($\kappa = 10^{-18}$).

## C. Optimization of a field with dissipation

We have also optimized a field in presence of dissipation [Eq.(27)] to mimic an experimental condition where dissipation should be active during feedback loops. The tested cases are $\kappa = 5$ $10^{-18}$ a.u. or $\bar{\gamma}^{-1} = 55$ $ms$ and $\kappa = 10^{-18}$ a.u. or $\bar{\gamma}^{-1} = 110$ $ms$. As expected, convergence is slow and the computation in density matrix is very time consuming. About 700 iterations require one month of CPU time on one processor (Intel(R) Xeon(R) CPU E5649 with frequency 2.53GHz, and 6 Go of RAM) and the fidelities obtained are 99.35% and 99.87% respectively. The Fourier transform of the field $E_p(t)$ calculated by Eq.(27) and the difference with the field calculated without dissipation (Fig. 4 (f)) are shown in Fig. 8. For the most favorable case $\kappa = 10^{-18}$ a.u. The control modifies more strongly the frequencies corresponding to $\Delta \nu = \pm 3$ and provides a spectrum with more background. Once more, the field amplitudes do not exceed 1.5 $Vm^{-1}$.

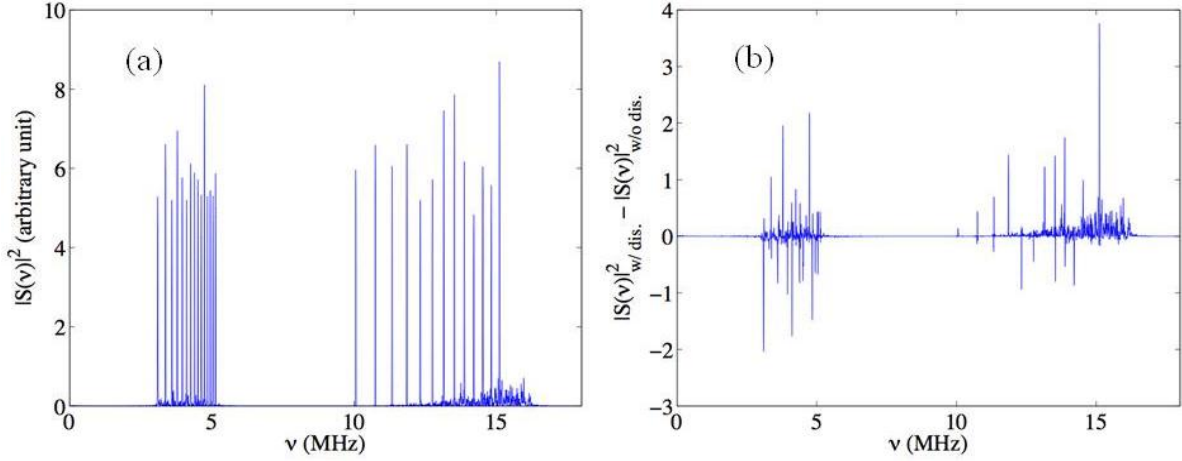

FIG. 8. Fourier transform $\left|S(\nu)\right|^2$ of the gate field $E_p(t)$ [Eq.(27)] optimized with dissipation (panel a) and the difference with the Fourier transform of the field optimized without dissipation (see Fig.4(f)) (panel b). The dissipation parameter is $\kappa = 10^{-18}$ a.u. or $\overline{\gamma}^{-1} = 110$ ms.

## V. CONCLUDING REMARKS

This work was stimulated by the recent experimental implementation of a TDSE simulator in NMR[72] and by the promising advancement of trapped-ion technologies. Following the numerical simulation of the Shor algorithm in an anharmonic trap by encoding information in the motional ionic states,[69] we have explored the same architecture to consider the TDSE simulator. This completes this previous work by two points. The TDSE simulation involves concatenation of several pulses and therefore a very strict control of the gate phase. Moreover, we check the robustness against decoherence. The TDSE unitary transformation corresponds to the evolution for a given time step. To calculate this unitary matrix, we adopt the Split Operator strategy among other possibilities whereas we use a "black box" approach for the optimization of the corresponding gate pulse. This means that we do not optimize a pulse for each elementary gate of the Split Operator algorithm since this would require a very large number of gates and therefore would increase the decoherence. Note that both phase shift transformations of the Split Operator sequence should be optimized for a given potential or a given mass. Only the quantum Fourier transform can be optimized once. Instead, we optimize a "black box" for a given time interval but this implies at least one computation of the matrix by any numerical algorithm. However, this gate remains identical for any initial condition $\psi(x, t = 0)$.

The fields obtained through optimal control remain very realistic. Their Fourier transform involve few frequencies in the *rf* domain. We have compared two strategies already discussed in the literature in order to ensure a good control of the gate phase.[24,89,90,92] Their convergence rate is similar and both lead to a very high fidelity. Only their spectrum is slightly different, indicating that different mechanisms are found by the control. This can affect the robustness against decoherence.

We also explored the decoherence time still allowing relevant results. In this case, it needs to be longer than about $\bar{\gamma}^{-1} = 110$ *ms* for a simulation of about 1 *ms*. We have used a crude model to simulate decoherence from heating. More sophisticated works should use the spectral density relative to a given architecture.[76,77,78,79] We could have taken other decoherence phenomena[58] into account, but this should not change the qualitative behavior observed in this work.

However, some qualitative insights on the dynamics can still be obtained even if the periodic behavior characteristic of the decoherence-free case is less and less preserved after sucessive applications of the pulse. The benchmark case of the harmonic potential should allow to test the decoherence since due to the very high fidelity of the optimum field, the phases at the end of the simulation are well described. One really obtain the prepared wave packet at a common arbitrary phase and the application of the reverse preparation field should give the ground motional state with a probability of one without dissipation. It is obvious that, in a general case, the goal is to get insights on the wave packet dynamics during its evolution and that the readout is a crucial step. Since the early works in this field, the manipulation of the motional states in order to create and read out different types of states have been examined.[95,96,97] Note that the target is mainly the localization probability of the evolving wave packet at the grids points since the control provides the unitary transformation with a global arbitrary phase so that the measure of the motional state population is sufficient. Such manipulation and population readout of the motional states of a single ion in a harmonic or anharmonic trap have been recently discussed in the context of verification of quantum thermodynamics.[98]

Finally, we have showed in this paper that the system is suitable to build a simulator of the elementary evolution operator for decoherence times larger than the simulation time by about a factor $10^2$. In order to construct an efficient and competitive simulator, it is obvious that scalability should be addressed. It is not conceivable to increase the number of qubits with a single ion. More dimensions[99] or more ions in the trap[100] in order to perform calculations with more points and thus more qubits have to be considered. The scheme involving only the two lowest motional states on each ion seems a promising perspective by manipulating cat states.[101,102]

**ACKNOWLEDGMENTS**

L.S acknowledges a F.R.I.A. research grant from the FRS-FNRS of Belgium. This work was supported by IISN contract n° 4.4504.10. We thank the European COST XLIC action and the French GDR THEMS.

[1] M. A. Nielsen and I. L. Chuang, *Quantum Computation and Quantum Information*, Cambridge U. P., Cambridge (2000).

[2] G. Benenti, G. Casati and G. Strini, *Principles of Quantum Computation and Information*, World Scientific, Singapore (2004).

[3] R. P. Feynman,Int. J. Theor. Phys. **21**, 467 (1982).

[4] I. Kassal, J. D. Whitfield, A. Perdomo-Ortiz, M.-H. Yung and A. Aspuru-Guzik, Annu. Rev. Phys. Chem. **62**, 185 (2011).

[5] A. Aspuru-Guzik, A. D. Dutoi, P. J. Love, and M. Head-Gordon, Science **309**, 1704 (2005).

[6] I. Kassal and A. Aspuru-Guzik, J. Chem. Phys. **131**, 224102 (2009).

[7] M.-H. Yung, J. Casanova, A. Mezzacapo, J. McClean, L. Lamata, A. Aspuru-Guzik and E. Solano, Scientific Reports| 4:3589| DOI: 10.1038/srep03589 (2014).

[8] R. Barends *et al*, arXiv:1501.07703v1 [quant-ph]

[9] A. Peruzzo, J. McClean, P. Shadbolt, M.-H. Yung, X. Q. Zhou, P. J. Love, A. Aspuru-Guzik and J. L. O'Brien, arxiv:1304.3061 (2013).

[10] I. Kassal, S. P. Jordan, P. J. Love, M. Mohseni and A. Aspuru-Guzik, PNAS **105**, 18681 (2008).

[11] A. T. Sornborer, Scientific Reports |2:597|, DOI: 10, 1038/srep000597 (2012).

[12] R. Gerritsma, G. Kirchmair, F. Zähringer, E. Solano, R. Blatt and C. F. Roos, Nature Lett. **463**, 68 (2010).

[13] W. Zhang, Sci. Bull. 60, 277 (2015).

[14] T. D. Ladd, F. Jelezko, R. Laflamme, Y. Nakamura, C. Monroe C and J. L. O'Brien, Nature **464**, 45 (2010).

[15] S. L. Braunstein and P. van Loock, Rev. Mod. Phys. **77**, 515 (2005).

[16] N. A. Gershenfeld and I. L. Chuang, Science **275**, 350 (1997).

[17] R. Hanson, L. P. Kouwenhoven, J. R. Petta, S. Tarucha and L. M. K. Vandersypen, Rev. Mod. Phys. **79**, 1217 (2007).

[18] Q. Morch and M. Oberthaler, Rev. Mod. Phys. **78**, 179 (2006).

[19] C. M. Tesch, L. Kurtz and R. de Vivie-Riedle, Chem. Phys. Lett. **243**, 633 (2001).

[20] C. M. Tesch and R. de Vivie-Riedle, Phys. Rev. Lett. **89**, 157901 (2002).

[21] J. Vala, Z. Amitay, B. Zhang B, S. Leone and R. Kosloff, Phys. Rev. A **66**, 62316 (2002).

[22] Z. Amitay, R. Kosloff, and S. R. Leone, Chem. Phys. Lett. **359**, 8 (2002).

[23] U. Troppmann, C. M. Tesch, and R. de Vivie-Riedle, Chem. Phys. Lett. **378**, 273 (2003).

[24] C. M. Tesch and R. de Vivie-Riedle, J. Chem. Phys. **121**, 12158 (2004).

[25] D. Babikov, J. Chem. Phys. **121**, 7577 (2004).

[26] B. Korff, U. Troppmann, K. Kompa and R. de Vivie-Riedle, J. Chem. Phys. **123**, 244509 (2005).

[27] U. Troppmann and R. de Vivie-Riedle, J. Chem. Phys. **122**, 154105 (2005).

[28] U. Troppmann, C. Gollub and R. de Vivie-Riedle, New J. Phys. **8**, 100 (2006).

[29] Y. Ohtsuki, Chem. Phys. Lett. **404**, 126 (2005).

[30] T. Cheng and A. Brown, J. Chem. Phys. **124**, 341115 (2006).

[31] M. Zhao and D. Babikov, J. Chem. Phys. **125**, 024105 (2006).

[32] D. Sugny, C. Kontz, M. Ndong, Y. Justum, G. Dive and M. Desouter-Lecomte, Phys. Rev. A **74**, 043419 (2006).

[33] M. Ndong, L. Bomble, D. Sugny, Y. Justum and M. Desouter-Lecomte, Phys. Rev. A **76**, 043424 (2007).

[34] M. Ndong, D. Lauvergnat, X. Chapuisat and M. Desouter-Lecomte, J. Chem. Phys. **126**, 244505 (2007).

[35] D. Weidinger and M. Gruebele, Mol. Phys. **105**,1999 (2007).

[36] M. Zhao and D. Babikov, J. Chem. Phys. **126**, 204102 (2007).

[37] M. Tsubouchi and T. Momose, Phys. Rev. A **77**, 052326 (2008).

[38] L. Bomble, D. Lauvergnat, F. Remacle, and M. Desouter-Lecomte, J. Chem. Phys. **128**, 064110 (2008).

[39] Y. Y. Gu and D. Babikov, J. Chem. Phys. **131**, 034306 (2009).

[40] R. R. Zaari and A. Brown, J. Chem. Phys. **132**, 014307 (2009).

[41] D. Sugny, L. Bomble, T. Ribeyre, O. Dulieu and M. Desouter-Lecomte, Phys. Rev. A **80**,042325 (2009).

[42] L. Bomble, D. Lauvergnat, F. Remacle, and M. Desouter-Lecomte, Phys. Rev. A **80**, 022332 (2009).

[43] L. Bomble, D. Lauvergnat, F. Remacle, and M. Desouter-Lecomte, Phys. Chem. Chem. Phys. **12**, 15628 (2010).

[44] Y. Ohtsuki, New J. Phys. **12**, 045002 (2010).

[45] K. Mishima and K. Yamashita, Chem. Phys. **376**, 63 (2010).

[46] R. R. Zaari and A. Brown, J. Chem. Phys. **135**, 044317 (2011).

[47] S. Sharma and H. Singh, Chem. Phys. **390**, 68 (2011).

[48] D. DeMille, Phys. Rev. Lett. **88,** 067901 (2002).

[49] L. D. Carr, D. DeMille, R. V. Krems and J. Ye, New J. Phys. **11**, 055049 (2009).

[50] S. F. Yelin, K. Kirby and R. Côté, Phys. Rev. A **74**, 050301 (2006).

[51] E. Kutznetsova, R. Côté, K. Kirby K and S. F. Yelin, Phys. Rev. A **78**, 012313 (2008).

[52] E. Charron, P. Milman, A. Keller and O. Atabek, Phys. Rev. A **78**, 012313 (2007).

[53] L. Bomble, P. Pellegrini, P. Ghesquière and M. Desouter-Lecomte, Phys. Rev. A **82**,062323 (2010).

[54] K. Mishima and K. Yamashita, Chem. Phys. **361**, 106 (2009).

[55] K. Mishima and K. Yamashita, J. Chem. Phys. **130**, 034108 (2009).

[56] Q. Wei, S. Kais, B. Friedrich and D. Herschbach, J. Chem. Phys. **135**, 154102 (2011).

[57] J. Zhu, S. Kais, Q. Wei, D. Herschbach and B. Friedrich, J. Chem. Phys. **138**, 024104 (2013).

[58] C. Monroe, R. Raussendorf, A. Ruthven, K. R. Brown, P. Maunz, L.-M. Duan and J. Kim, Phys. Rev. A **89**, 022317 (2014).

[59] J. I. Cirac and P. Zoller, Phys. Rev. Lett**. 74**, 4091 (1995).

[60] D. J. Wineland, C. Monroe, W. M. Itano, D. Leibfried, B. E. King, and D. M. Meekhof, J. Res. Natl. Inst. Stand. Technol. **103**, 259 (1998).

[61] C. Monroe, D. M. Meekhof, B. E. King, W. M. Itano and D. J. Wineland, Phys. Rev. Lett. **75**, 4714 (1995).

[62] L-M. Duan and C. Monroe, Rev. Mod. Phys. **82**, 1209 (2010).

[63] C. Monroe and J. Kim, Science **339**, 164 (2013).

[64] T. P. Harty, D.T.C. Allcock, C.J. Balance, L. Guidoni, H. A. Janachek, N. M. Linke, D. N. Stacey, and D. M. Lucas, Phys. Rev. Lett. 113, 220501 (2014).

[65] H. Häffner, C. F. Roos and R. Blatt, Phys. Rep. **469**, 155 (2008).

[66] G.-D. Lin, S.-L. Zhu, R. Islam, K. Kim, M.-S. Chang, S. Korenblit, C. Monroe and L.-M. Duan, Europhys. Lett. **86**, 60004 (2009).

[67] M. Zhao and D. Babikov, Phys. Rev. A **77**, 012338 (2008).

[68] L. Wang and D. Babikov, Phys. Rev. A **83**, 022305 (2011).

[69] L. Wang and D. Babikov, J. Chem. Phys. **137**, 064301 (2012).

[70] Q. Chen, K. Hai and W. Hai, J. Phys. A Math. Theor. **43**, 455302 (2010).

[71] G. Benenti and G. Strini, Am. J. Phys. **76**, 657 (2008).

[72] M. D. Feit, Jr J. A. Fleck, and A. Steiger, J. Comput. Phys. **47**, 412 (1982).

[72] D. Lu, N. Xu, R. Xu, H. Chen, J. Gong, X. Peng and J. Du, Phys. Rev. Lett. **107**, 020501 (2011).

[73] D. Lu, B. Xu, N. Xu, Z. Li, H. Chen,  X. Peng, R. Xu and J. Du, Phys. Chem. Chem. Phys. **14**, 9411 (2012).

[74] H. Tal-Ezer and R. Kosloff, J. Chem. Phys. **81**, 3967 (1984).

[75] S. Schneider and G. J. Milburn, Phys. Rev. A **59**, 3766 (1999).

[76] Q. A. Turchette, D. Kielpinski, B. E. King, D. Leibfried, D. M. Meekhof, C. J. Myatt, M. A. Rowe, C. A. Sackett, C. S. Wood, W. M. Itano, C. Monroe and D. J. Wineland, Phys. Rev. A **61**, 063418 (2000).

[77] A. Safavi-Naini, P. Rabl, P. F. Weck and H. R. Sadeghpour, Phys. Rev. A **84**, 023412 (2011).

[78] A. Safavi-Naini, E. Kim, P. F. Weck, P. Rabl and H. R. Sadeghpour, Phys. Rev. A **87**, 023421 (2011).

[79] N. Daniilidis, S. Narayanan, S. A. Möller, R. Clarck, T. E. Lee, P. J. Leek, A. Wallraff, St Schulz, F. Schmidt-Kaler and H. Häffner, New J. Phys. **14**, 079504 (2012).

[80] B. B. Blinov, D. L. Moehring, L.-M. Duan and C. Monroe, Nature **428**, 153 (2004).

[81] L. Deslauriers, S. Olmschenk, D. Stick, W. Heisinger, J. Sterk, and C. Monroe, Phys. Rev. Lett. **97**, 103007 (2006).

[82]  See Supplementary Material Document No.__________ for the eigenenergies and for the dipole matrix.

[83] G. Lindblad, Commun. Math. Phys. **48**, 119 (1976).

[84] W. Zhu and H. Rabitz, J. Chem. Phys. **118**, 6751 (2003).

[85] F. Shuang and H. Rabitz, J. Chem. Phys. **124**, 154105 (2006).

[86] H. F. Trotter, Proc. Amer. Math. Soc. **10**, 545 (1959).

[87] J. L. Walsh, Am. J. Math. **45**, 5 (1923).

[88] J. Welch, D. Greenbaum, S. Mostame and A. Aspuru-Guzik, New J. Phys. **16**, 033040 (2014).

[89] J. P. Palao and R. Kosloff, Phys. Rev. Lett. **89,** 188301 (2002).

[90] J. P. Palao and R. Kosloff, Phys. Rev. A **68,** 062308 (2003).

[91] W. Zhu, J. Botina and H. Rabitz, J. Chem. Phys. **108**, 1953 (1998).

[92] A. Jaouadi, E. Barrez, Y. Justum and M. Desouter-Lecomte, J. Chem. Phys. **139**, 014310 (2013).

[93] Y. Ohtsuki, W. Zhu and H. Rabitz, J. Chem. Phys. **110**, 9825 (1999).

[94] W. H Press, S. A. Teukolsky, W. T. Vetterling, and B. P. Flannery, *The Art of Scientific Computing,* Cambridge University Press, Cambridge, (2007).

[95] D. Leibfried, D.M. Meekhof, B. E. King, C. Monroe, W. M. Itano, and D.J. Wineland, Phys. Rev. Lett. **77**, 4281 (1996).

[96] D. M. Meekhof, C. Monroe, B. E. King, W. M. Itano, and D. J. Wineland, Phys. Rev. Lett. **76**, 1796 (1996).

[97] J. C. Retanal and Z. Zagury, Phys. Rev. A **55**, 2387 (1997).

[98] G. Huber, F. Schmidt-Kaler, S. Deffner, and E. Lutz, Phys. Rev. Lett. 101, 070403 (2008).

[99] X.-B. Zou, J. Kim, and H.-W. Lee, Phys. Rev. A **63**, 065801 (2001).

[100] E. Solano, P. Milman, R.L. de Matos Filho, and Z. Zagury, Phys. Rev. A **62**, 021401(R) (2000).

[101] H. Häffner *et al*. Nature, 438, 643 (2005).

[102] D. Liebfried *et al*. Nature, 438, 639 (2005).